\documentclass[letterpaper, 10pt, conference]{ieeeconf}      % Use this line for a4 paper
\renewcommand{\baselinestretch}{0.925}

\IEEEoverridecommandlockouts                             % This command is only
\usepackage[usenames,dvipsnames]{xcolor}
\usepackage{graphicx} % for pdf, bitmapped graphics files
\usepackage{mathptmx} % assumes new font selection scheme installed
\usepackage{times} % assumes new font selection scheme installed
\usepackage{amsmath} % assumes amsmath package installed
\usepackage{amssymb}  % assumes amsmath package installed
\usepackage{pstricks,pst-plot,psfrag}
\usepackage{units}
\usepackage{multirow}
\usepackage[draft]{hyperref}
\usepackage[]{mcode}
\usepackage{booktabs}
\usepackage[acronym]{glossaries}
\usepackage{cite}
\usepackage{lettrine}
\usepackage{import}
\usepackage{tikz}
\usetikzlibrary{positioning}
\usetikzlibrary{calc}

\definecolor{lightblue}{rgb}{0.60784,0.76078,0.90196}
\definecolor{darkblue}{rgb}{0.26667,0.44706,0.76863}
\definecolor{lightgreen}{rgb}{0.66275,0.81569,0.55686}
\definecolor{darkgreen}{rgb}{0.43922,0.67843,0.27843}
\definecolor{orange}{rgb}{0.92941,0.49020,0.19216}
\definecolor{yellow}{rgb}{1.00000,0.75294,0.00000}
\definecolor{grey}{rgb}{0.64706,0.64706,0.64706}
\definecolor{purple}{rgb}{0.51373,0.23529,0.04706}
\newacronym{abk:amod}{AMoD}{Autonomous Mobility-on-Demand}
\newacronym{abk:iamod}{I-AMoD}{intermodal \gls{abk:amod}}
\newacronym{abk:ffcs}{FFCS}{free floating car sharing systems}
\newacronym{abk:ghg}{GHG}{greenhouse gas}
\newacronym{abk:mcfp}{MCFP}{multi-commodity flow problem}
\newacronym{abk:spp}{SPP}{shortest path problem}
\newacronym{abk:kdspp}{k-dSPP}{k-disjoint \gls{abk:spp}}
\newcommand{\vTime}{V_\mathrm{T}}
\newcommand{\vDistR}{V_\mathrm{D,R}}
\newcommand{\vDistS}{V_\mathrm{D,P}}
\newcommand{\vEnergy}{V_\mathrm{E}}

\newcommand{\vQuadratic}{V_\mathrm{Q}}
\newcommand{\capacityRoad}{c^{\mathrm{R}}}
\newcommand{\capacityPublic}{c^{\mathrm{P}}}
\newcommand{\cost}{J}
\newcommand{\costM}{J_\mathrm{M}}

\newcommand{\destination}{d}

\newcommand{\energyR}{e_{\mathrm{R},ij}}

\newcommand{\flow}[2]{f_m\left(#1,#2\right)}
\newcommand{\flowzero}[2]{f_m\left(#1,#2\right)}
\newcommand{\flowReba}[2]{f_0\left(#1,#2\right)}
\newcommand{\origin}{o}
\newcommand{\pSubway}{p_\mathrm{P}(i,j)}
\newcommand{\pRoad}{p_\mathrm{R}(i,j)}
\newcommand{\pOrigin}[1]{p_\mathrm{O}\left(#1\right)}
\newcommand{\pDest}[1]{p_\mathrm{D}\left(#1\right)}
\newcommand{\request}{r}

\newcommand{\traveltime}{t}
\newcommand{\tRoad}{\tau_\mathrm{R}(i,j)}
\newcommand{\setOfArcs}{\mathcal{A}}
\newcommand{\setOfArcsRoad}{\mathcal{A}_{\mathrm{R}}}
\newcommand{\setOfArcsSubway}{\mathcal{A}_{\mathrm{P}}}
\newcommand{\setOfArcsPedestrian}{\mathcal{A}_{\mathrm{W}}}
\newcommand{\setOfArcsCommute}{\mathcal{A}_{\mathrm{C}}}
\newcommand{\GraphRoad}{\mathcal{G}_\mathrm{R}}
\newcommand{\GraphSubway}{\mathcal{G}_\mathrm{P}}
\newcommand{\GraphPedestrian}{\mathcal{G}_\mathrm{W}}
\newcommand{\setOfRequestsNumber}{\mathcal{M}}
\newcommand{\setOfRequests}{\mathcal{R}}
\newcommand{\setOfVertices}{\mathcal{V}}
\newcommand{\setOfVerticesRoad}{\mathcal{V}_{\mathrm{R}}}
\newcommand{\setOfVerticesSubway}{\mathcal{V}_{\mathrm{P}}}
\newcommand{\setOfVerticesPedestrian}{\mathcal{V}_{\mathrm{W}}}
\newcommand{\requestrate}{\alpha}
\newcommand{\dualCust}{\lambda_\mathrm{C}}
\newcommand{\dualVeh}{\lambda_\mathrm{R}}
\newcommand{\dualCustTilde}{\tilde{\lambda}_\mathrm{C}}
\newcommand{\dualVehTilde}{\tilde{\lambda}_\mathrm{R}}
\newcommand{\dualcR}{\mu_\mathrm{cR}}
\newcommand{\dualcS}{\mu_\mathrm{cP}}
\newcommand{\arc}{(i,j)}

\newcommand{\bool}[1]{\mathds{1}_{#1}}

\newcommand{\cG}{\mathcal{G}}

\newcommand{\cM}{\mathcal{M}}

\newcommand{\cV}{\mathcal{V}}

\newcommand{\sR}{\mathbb{R}}
\usepackage{mathtools}
\usepackage{dsfont}
\usepackage{booktabs}
\newtheorem{theorem}{Theorem}[section]

\title{\LARGE \bf
On the Interaction between Autonomous Mobility-on-Demand\\ and Public Transportation Systems
}

\author{Mauro Salazar$^{1,2}$, Federico Rossi$^2$, Maximilian Schiffer$^{2,3}$, Christopher H.\ Onder$^1$, and Marco Pavone$^2$
\thanks{$^1$Institute for Dynamic Systems and Control
        ETH Z\"urich, Zurich (ZH), Switzerland
        {\tt maurosalazar@idsc.mavt.ethz.ch}}%
\thanks{$^2$Autonomous Systems Lab, Stanford University, Stanford (CA), United States
        {\tt \{frossi2,pavone\}@stanford.edu}}%
\thanks{$^3$TUM School of Management, Technical University of Munich, 80333 Munich, Germany
   	{\tt maximilian.schiffer@tum.de}}
}
\newcommand{\summe}[1]{\sum_{\scriptstyle\mathclap{#1}}}

\newif\ifmargincomments %A quick way of turning off margin comments for, say, arXiv submission
\margincommentstrue

\newif\ifextendedversion %A quick way of turning off appendix
\extendedversiontrue

\ifmargincomments

\else

\fi

\begin{document}
\maketitle
\thispagestyle{plain}
\pagestyle{plain}

%%%%%%%%%%%%%%%%%%%%%%%%%%%%%%%%%%%%%%%%%%%%%%%%%%%%%%%%%%%%%%%%%%%%%%%%%%%%%%%%
\begin{abstract}
In this paper we study models and coordination policies for intermodal \gls{abk:amod}, wherein a fleet of self-driving vehicles provides on-demand mobility jointly with public transit.
Specifically, we first present a network flow model for intermodal \gls{abk:amod}, where we capture the coupling between \gls{abk:amod} and public transit and the goal is to maximize social welfare. Second, leveraging such a model, we design a pricing and tolling scheme that allows to achieve the social optimum under the assumption of a perfect market with selfish agents. Finally, we present a real-world case study for New York City. Our results show that the coordination between \gls{abk:amod} fleets and public transit can yield significant benefits compared to an \gls{abk:amod} system operating in isolation.
\end{abstract}

%%%%%%%%%%%%%%%%%%%%%%%%%%%%%%%%%%%%%%%%%%%%%%%%%%%%%%%%%%%%%%%%%%%%%%%%%%%%%%%%
%% !TEX root = ../paper.tex
\section{Introduction}\label{sec:introduction}
\lettrine{T}{raffic} congestion is soaring all around the world. Besides mere discomfort for passengers, congestion causes severe economic and environmental harm, e.g., due to the loss of working hours and pollutant emissions such as  CO\textsubscript{2}, particulate matter, and NO\textsubscript{x}~\cite{LevyBuonocoreEtAl2010}. In 2013, traffic congestion cost U.S. citizens \$124 Billion~\cite{TuttleCowles2014}. Notably, transportation remains one of a few sectors in which emissions are still increasing~\cite{EPA2018}. Governments and municipalities are struggling to find sustainable ways of transportation that can match mobility needs and reduce environmental harm as well as congestion.

To achieve sustainable modes of transportation, new mobility concepts and technology changes are necessary. However, the potential to realize such concepts in urban environments is limited, since the available infrastructures (e.g., roads and subway lines) and their capacity are given and mostly fixed.
Thus, mobility concepts that use the existing infrastructure in a more efficient way are sought. 
Concepts focusing on mobility-on-demand services are particularly promising.
Herein, two main concepts exist.
On the one hand, free floating car sharing systems strive to reduce the number of private cars in city centers. However, these systems offer limited flexibility and are generally characterized by low adoption rates.
One reason for this is the low vehicle availability due to the difficulty of \emph{rebalancing} empty vehicles to counter asymmetric customer demand. On the other hand, ride-hailing systems aim to enhance and extend the service of taxi fleets. However, current studies show that ride-hailing services can worsen traffic congestion significantly due to the heightened demand and vehicle-miles traveled by customer-empty vehicles. Additionally, the low cost and the point-to-point nature of ride-hailing services may lead both to an induced demand and a demand shift from other modes of transit, as is currently the case for Manhattan. Indeed, recent studies for the Manhattan area revealed the massive magnitude of this issue: From 2013 to 2018, the number of for-hire vehicles exploded from 47'000 to 103'000, 68'000 of which are employed for ride-hailing services. Due to this increase, the average traffic speed dropped by 13\% from \unit[6.5]{mph} to \unit[4.7]{mph}~\cite{Hu2017}.
\begin{figure}[t]
	\centering\includegraphics[width=\columnwidth]{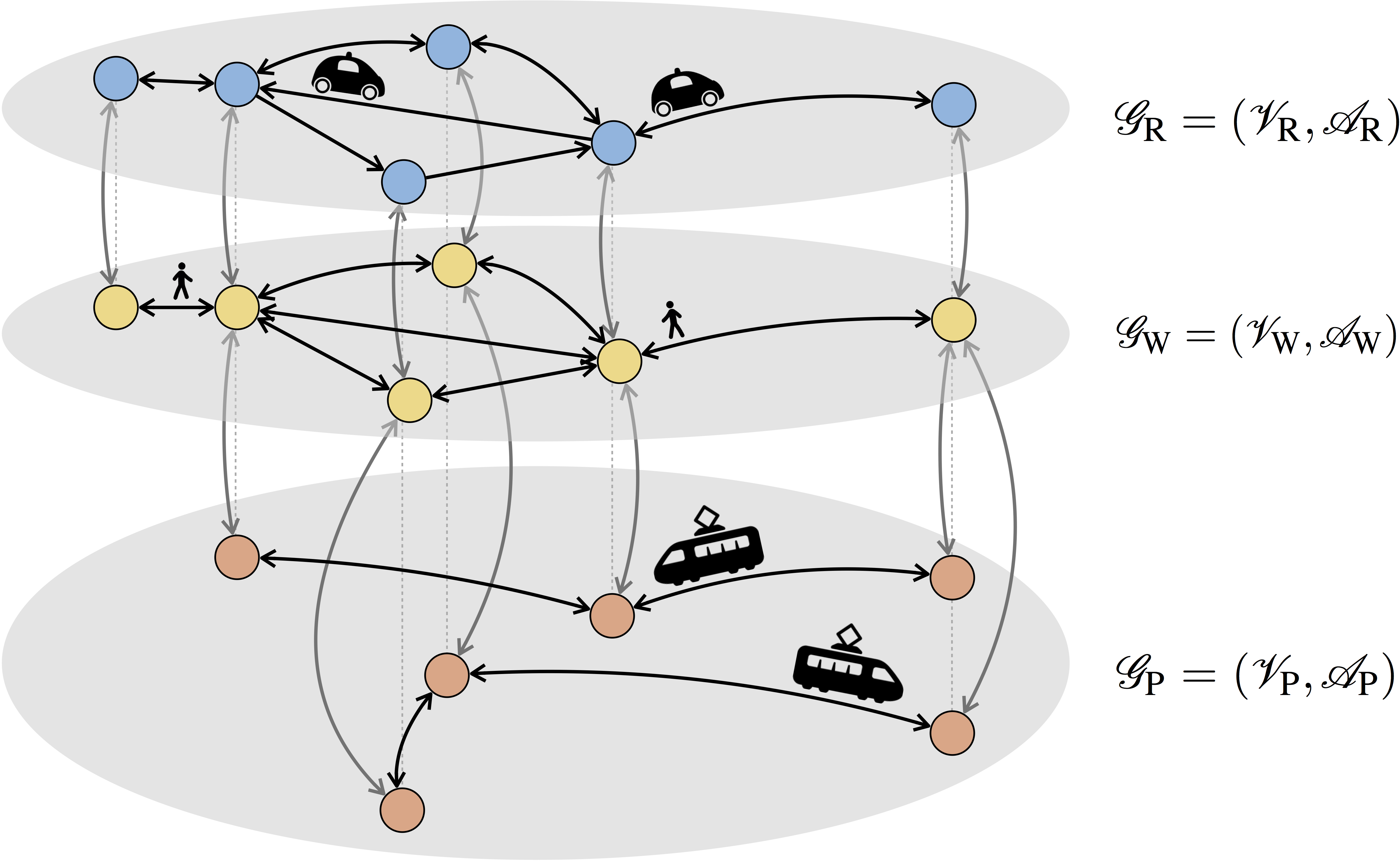}
	\caption{The intermodal AMoD network consists of a road digraph, a walking digraph and a public transportation digraph. The colored dots denote intersections or stops and the black arrows represent road links, pedestrian pathways or public transit arcs.
		The grey dotted lines denote geographically close nodes while the grey arrows are the mode-switching arcs that connect them.}
	\label{fig:digraph}
	\vspace{-5mm}
\end{figure}

Autonomous Mobility-on-Demand (\gls{abk:amod}) systems hold promise as a future mobility concept in urban environments.
\gls{abk:amod} systems comprise a fleet of robotic, self-driving vehicles that transport passengers between their origins and destinations. A central operator runs such systems by assigning passenger requests to vehicles and computing rebalancing routes for the unassigned empty vehicles, in order to re-align their geographical distribution with the transportation demand. Thus an \gls{abk:amod} system can replace a conventional taxi, car sharing, or ride-hailing fleet, while offering several advantages compared to the previously discussed concepts: First, no relocation costs for drivers arise; second, much higher vehicle utilization rates can be achieved compared to car-sharing systems thanks to continuous rebalancing; third, the centralized control of the complete fleet allows for more operational flexibility and efficiency compared to ride-hailing or taxi fleets where a central operator can suggest, but not directly control, vehicle routes, and enables the operator to adopt fleet-wide routing strategies to mitigate congestion.
However, operating an \gls{abk:amod} system to cover the complete transportation demand of a city would inexorably increase the number of operated vehicles and cause congestion again due to induced demand for transportation, as customers shift from public transit to shared cars.
For an \gls{abk:amod} system to be efficiently deployed, it must intelligently cooperate with other modes of transportation, such as the public transportation network, in order to secure sustainable and congestion-free urban mobility. Against this backdrop, our study aims to develop modeling and optimization tools to assess and realize the benefits of the intermodal transportation system shown in Fig.~\ref{fig:digraph} that combines public transit with \gls{abk:amod}.

{\em Related literature}: Our work contributes to three different research streams, namely: i) \gls{abk:amod} systems, ii) congestion pricing, and iii) multimodal passenger transport. In the following, we review these research streams in turn.

A number of approaches to characterize and control AMoD systems in isolation are available, ranging from queuing-theoretical models \cite{ZhangPavone2016} to simulation-based models \cite{LevinKockelmanEtAl2017,MaciejewskiBischoffEtAl2017,HoerlRuchEtAl2018} and multi-commodity network flow models \cite{RossiZhangEtAl2017,SpieserTreleavenEtAl2014}. Queueing-theoretical models capture the stochasticity of the customer arrival process and are amenable to efficient control synthesis. However, their complex structure makes it difficult to capture the interaction with other modes of transportation. Simulation-based models capture transportation systems with very high fidelity but are generally not amenable to optimization. 
Network flow models are amenable to efficient optimization and allow for the inclusion of a variety of complex constraints. Accordingly, they have seen wide use in problems ranging from control of AMoD systems in congested road networks \cite{RossiZhangEtAl2017}, to cooperative control of AMoD systems and the electric power network \cite{RossiIglesiasEtAl2018}, and control of human-driven MoD systems \cite{PavoneSmithEtAl2012}.

Congestion pricing in general has been widely investigated, for instance by analyzing Pigovian taxes for network congestion problems \cite{MayeresProost1996}, but only few approaches focus on pricing in the context of \gls{abk:amod}: Specifically, \cite{SimoniKockelmanEtAl2018} focuses on congestion pricing for self-driving vehicles by incentivizing socially and environmentally aware travel modes, while \cite{ChenKockelman2016} proposes pricing schemes to foster the use of \gls{abk:amod} systems.
However, these studies comprise logit modeling approaches and rely on agent-based simulations. As such, they assess the performance of \emph{pre-determined} \gls{abk:iamod} routing policies. In contrast, our optimization-based approach identifies the \emph{best} achievable performance of an \gls{abk:iamod} system and enables the synthesis of policies that steer a system towards such an optimum.

Literature on intermodal passenger transportation is still sparse. First, studies on the interplay between \gls{abk:amod} and public transportation exist, focusing either on fluidic~\cite{VakayilGruelEtAl2017} or simulation-based~\cite{GentileNoekel2016,BischoffKaddouraEtAl2017,MaciejewskiBischoffEtAl2017} models.
However, these studies focus on the analysis of specific scenarios, as opposed to the \emph{optimization} of joint control policies for AMoD systems and public transit.
In general, to the best of the authors' knowledge, only descriptive analyses of intermodal passenger transport exist~\cite{SeabornAttanucciEtAl2009}.

In summary, some optimization approaches and control policies for \gls{abk:amod} systems are available but lack the consideration of public transit. With respect to pricing schemes, existing studies develop pricing schemes to address individual externalities (e.g., congestion), but not a single study captures the interplay between multiple externalities arising from the synchronization of different modes of transportation. Finally, there exist no optimization frameworks that are capable of determining optimal coordination policies for \gls{abk:iamod} systems and assessing their achievable performance. In this paper, by coordination we mean that customer routes are optimized by jointly accounting for \gls{abk:amod} services and public transportation services.

{\em Statement of contributions}:  The goal of this paper is to introduce a mesoscopic optimization approach for \gls{abk:iamod} systems. Specifically, the contribution of this paper is fourfold: First, we develop a multi-commodity network flow optimization model that captures the joint operations of \gls{abk:amod} systems and public transit.
In our model, the objective is to maximize the social welfare, i.e., to minimize the customers' travel time together with the operational costs of different transportation modes. Herein, we also consider congestion effects.
Second, we propose a pricing and tolling scheme that helps to realize the social optimum in the presence of selfish actors such as customers and \gls{abk:amod} operators. Third, we present a real-world case study for Manhattan. Fourth, we derive managerial insights on the benefits of \gls{abk:iamod} systems: Our results show that an \gls{abk:iamod} system can significantly reduce travel times, pollutant emissions, total number of cars, and overall costs with respect to an \gls{abk:amod} system operating in isolation (i.e., without coordination with public transit). Interestingly, the resulting pricing and tolling scheme is aligned with the recently proposed congestion surcharges for ride-hailing vehicles.

{\em Organization}: The remainder of this paper is structured as follows: In Section~\ref{sec:model} we present the flow optimization model for \gls{abk:iamod}. Section~\ref{sec:pricing} derives a pricing and tolling scheme to steer self-interested agents toward the social optimum. Section~\ref{sec:results} presents the case study of Manhattan. Finally, Section~\ref{sec:conclusion} concludes the paper with a short summary and an outlook on future research.
%% !TEX root = ../Salazar.Rossi.ea.ITSC18.tex
\section{Flow Optimization Model of \gls{abk:iamod}}\label{sec:model}
This section introduces a flow optimization approach for intermodal \gls{abk:amod} systems. In this approach, we consider i) the assignment of transportation requests to transport flows, ii) different modes of transportation, iii) capacity limits which are specific to the transportation mode,
and iv) rebalancing activities to re-align the distribution of empty vehicles with transportation demand.
Assuming a centrally controlled system, Section~\ref{subsec:model} introduces a generic multi-commodity flow based optimization approach, while Section~\ref{subsec:objective} details its objective and Section~\ref{sec:model:discussion} discusses our assumptions.

\subsection{Multi-commodity Flow Based Optimization Approach}\label{subsec:model}
To represent the transportation system and its different transportation modes, we use the digraph $\cG=(\setOfVertices,\setOfArcs)$ shown in Fig.~\ref{fig:digraph}, which has a set of vertices $\setOfVertices$ and a set of arcs $\setOfArcs\subseteq\setOfVertices\times\setOfVertices$. The graph contains a road network layer $\GraphRoad = (\setOfVerticesRoad,\setOfArcsRoad)$, a public transportation layer $\GraphSubway=(\setOfVerticesSubway,\setOfArcsSubway)$, and a walking layer $\GraphPedestrian=(\setOfVerticesPedestrian,\setOfArcsPedestrian)$. Herein, the road layer represents intersections $i\in\setOfVerticesRoad$ and road links $\arc\in\setOfArcsRoad$. The public transportation layer comprises stops $i\in\setOfVerticesSubway$ connected by arcs $\arc\in\setOfArcsSubway$, while the pedestrian layer represents walkable streets $\arc\in\setOfArcsPedestrian$ between intersections $i\in\setOfVerticesPedestrian$. Finally, mode-switching arcs out of set $\setOfArcsCommute\subseteq\setOfVerticesRoad\times\setOfVerticesPedestrian \cup \setOfVerticesSubway\times\setOfVerticesPedestrian$ connect the pedestrian layer to the road and public transportation layers, respectively, whereby $\setOfVerticesRoad\cap\setOfVerticesSubway=\emptyset$. These arcs model the customer's ability to switch transportation modes, for instance by exiting a vehicle, taking the subway or hailing an \gls{abk:amod} ride. Collecting all definitions, it holds that $\setOfVertices = \setOfVerticesPedestrian\cup \setOfVerticesRoad \cup \setOfVerticesSubway$ and $\setOfArcs = \setOfArcsPedestrian \cup \setOfArcsRoad \cup \setOfArcsSubway \cup \setOfArcsCommute$. 

Traversing an arc $\arc$ takes on average $\traveltime_{ij}$ time units.
For mode-switching arcs such a parameter denotes the time necessary to switch between two means of transportation, such as exiting a vehicle, waiting for the bus or hailing an \gls{abk:amod} ride.
To describe congestion we use a simple threshold model: We denote by $\capacityRoad_{ij}$ the capacity of each road arc. When the flow-rate on a road arc is less than the capacity of that arc, all vehicles are assumed to travel at free-flow speed, with a corresponding free-flow traversal time given by $\traveltime_{ij}$. Conversely, when the flow-rate is larger than the capacity of the road arc, the traversal time is set equal to infinity.
Similarly, we denote by $\capacityPublic_{ij}$ the capacity of public transportation arcs. When the flow-rate on a public transportation arc is less than the capacity of that arc, all passengers are assumed to travel with a free-flow traversal time given by $\traveltime_{ij}$.
Otherwise, the traversal time is set equal to infinity.
The customer flows on walking or mode-switching arcs are not constrained, as we assume enough space on sidewalks and in public transit stations.

Let $\setOfRequests$ be the set of all travel requests. Rigorously, a request $\request_m = \left(\origin_m,\destination_m,\requestrate_m\right)\in\setOfRequests$ is a triple comprising an origin node $\origin_m\in\setOfVerticesPedestrian$, a destination node $\destination_m\in\setOfVerticesPedestrian$, and a request rate $\requestrate_m$ that denotes the amount of customers per unit time for each request $m\in\setOfRequestsNumber = \{1,\dots,M\}$. Note that $\origin_m$ and $\destination_m$ lie on the walking digraph. Considering the different transportation modes, $\flow{i}{j}$ denotes the flow (i.e., the number of customers per unit time) on arc $\arc\in\setOfArcs$ for a certain travel request $m$. To account for rebalancing flows between a customer's destination and the next customer's origin, $\flowReba{i}{j}$ denotes the flow of empty vehicles on road arcs $\arc\in\setOfArcsRoad$.

For a given cost function $\cost$ that maps the set of flows $\{\flow{\cdot}{\cdot}\}_m,\flowReba{\cdot}{\cdot}$ into the set of non-negative real numbers $\sR_{\geq0}$, the \gls{abk:iamod} optimization problem can be stated as

\begingroup
\allowdisplaybreaks
\begin{small}
	\begin{subequations}
		\label{eq:IAMoD}
		\begin{align}
		&\min_{\{\flow{\cdot}{\cdot}\}_m,\flowReba{\cdot}{\cdot}} \quad\cost\big(\{\flow{\cdot}{\cdot}\}_m,\flowReba{\cdot}{\cdot}\big)
		\label{eq:obj}\\
		&\text{s.t.}\nonumber \\
		&\summe{i:\arc\in\setOfArcs}f_m(i,j) + \bool{j=o_m}\cdot \alpha_m = \summe{k:(j,k)\in\setOfArcs}f_m(j,k) + \bool{j=d_m}\cdot \alpha_m \span\nonumber\\
		& & \forall m\in\cM,\, j\in\cV\label{eq:customerbalance}\\
		&\sum_{i:\arc\in\setOfArcsRoad}\left(\flowReba{i}{j} + \summe{m\in\setOfRequestsNumber}f_m(i,j)\right) = \nonumber\\
		&\quad  \sum_{k:(j,k)\in\setOfArcsRoad} \left((\flowReba{j}{k} + \summe{m\in\setOfRequestsNumber}f_m(j,k)\right) & \forall j\in\setOfVerticesRoad
		\label{eq:flowreba}\\
		&\flowReba{i}{j} + \summe{m\in\setOfRequestsNumber}\flowzero{i}{j} \leq \capacityRoad_{ij} & \forall\arc\in\setOfArcsRoad
		\label{eq:caproads}\\
		&\summe{m\in\setOfRequestsNumber}\flow{i}{j} \leq \capacityPublic_{ij} &\forall\arc\in\setOfArcsSubway.
		\label{eq:cappublic}
		\end{align}
	\end{subequations}%
\end{small}
\endgroup
\!\!\!\!
For a given set of transportation demands $(\origin_m,\destination_m,\requestrate_m)\in\setOfRequests$, we compute the optimal customer flows $\{\flow{\cdot}{\cdot}\}_m$ and rebalancing flows $\flowReba{\cdot}{\cdot}$ minimizing the cost $\cost$ in Eq.~\eqref{eq:obj}. The constraint~\eqref{eq:customerbalance} guarantees flow conservation for customers, where $\bool{j=x}$ is a boolean indicator function. We enforce flow conservation for vehicles in Eq.~\eqref{eq:flowreba}, and capacity limits for road links in Eq.~\eqref{eq:caproads} and public transportation links in Eq.~\eqref{eq:cappublic}.

\subsection{\gls{abk:iamod} Objective}\label{subsec:objective}
The general cost function~\eqref{eq:obj} can be used to capture different objectives. In this paper, we are interested in optimizing the social welfare by minimizing the passengers' travel time and the operational costs incurred by the \gls{abk:iamod} system. Specifically, we define commuting costs that depend on the customers' value of time and on operational costs for the \gls{abk:amod} fleet and the public transportation. Herein, we assume the same value of time $\vTime$ for each customer, while costs for the \gls{abk:amod} fleet comprise distance-dependent ownership costs $\vDistR$ to account for maintenance and depreciation as well as energy costs $\vEnergy$. For the public transit system, we condense all operational costs per passenger kilometer in the parameter $\vDistS$. Finally, we consider a quadratic regularization term with a very small weight $\vQuadratic$. While this regularization term does not appreciably influence the overall cost, it does ensure strict convexity for the problem and thus uniqueness of a solution -- a key property that will be leveraged in Section~\ref{sec:pricing} to design a socially-optimal pricing and tolling scheme. The social cost is then defined as
\begin{equation}
\begin{aligned}\label{eq:costM}
\costM&\big(\{\flow{\cdot}{\cdot}\}_m,\flowReba{\cdot}{\cdot}\big) = \vTime\cdot \summe{m\in\setOfRequestsNumber,\arc\in\setOfArcs}\traveltime_{ij}\cdot \flow{i}{j}\\
+&\summe{\arc\in\setOfArcsRoad}(\vDistR\cdot  d_{ij}+\vEnergy\cdot \energyR) \cdot \Bigg(\flowReba{i}{j}+\summe{m\in\setOfRequestsNumber}\flow{i}{j}\Bigg)\\
+&\vDistS\cdot \summe{\arc\in\setOfArcsSubway}d_{ij}\cdot \summe{m\in\setOfRequestsNumber}\flow{i}{j}\\
+&\vQuadratic\cdot\left( \sum_{m\in\setOfRequestsNumber}\sum_{\arc\in\setOfArcs}\flow{j}{j}^2 + \sum_{\arc \in \setOfArcsRoad}\flowReba{i}{j}^2\right).\hfill
\end{aligned}
\end{equation}
Given the mesoscopic nature of our study, we estimate the energy consumption of a single vehicle $\energyR>0,\,\arc\in\setOfArcsRoad$ assuming that road arcs are traversed at the constant speed $v_{ij}=d_{ij}/t_{ij}$. 
Considering electric vehicles with full recuperation capabilities and an overall tank-to-wheel efficiency $\eta_\mathrm{EV}$, the energy consumption for a road arc is
\begin{equation}\label{eq:energyR}
\energyR = \left(\frac{\rho_\mathrm{a}}{2} \cdot A_\mathrm{f} \cdot c_\mathrm{d} \cdot v_{ij}^2+c_\mathrm{r}\cdot m_\mathrm{v}\cdot g\right)\cdot \frac{ d_{ij}}{\eta_\mathrm{EV}}  \quad\!\! \forall \arc\in\setOfArcsRoad.
\end{equation}
The first term in \eqref{eq:energyR} represents the aerodynamic drag composed by the air density $\rho_\mathrm{a}$, the frontal area $A_\mathrm{f}$, the drag coefficient $c_\mathrm{d}$, and the rolling friction computed combining its coefficient $c_\mathrm{r}$ with the mass of the vehicle $m_\mathrm{v}$ and the gravitational acceleration $g$ \cite{GuzzellaSciarretta2007}.

\subsection{Discussion}\label{sec:model:discussion}
A few comments are in order.
First, we consider time-invariant travel requests. This assumption is valid if requests change slowly compared to the average travel time of an individual trip, as is often the case in densely populated urban environments~\cite{Neuburger1971}. Second, we adopt a simple threshold model for congestion. The model is consistent with classical traffic flow theory \cite{Wardrop1952} and it is adequate for the goal of efficiently optimizing customer and vehicle routes. Congestion models offering higher accuracy can be used for the \emph{analysis} of specific control policies. Third, the model in this paper represents customer and vehicle routes as fractional flows and does not capture the stochastic nature of the customer arrival process and the residual traffic in the network. These approximations are in line with the mesoscopic nature of our study. For real-time control applications, randomized routing algorithms can be employed to compute integer-valued flows starting from the fractional flows that yield near-optimal routes for individual customers\cite[Ch. 4]{Rossi2018} and new information can be accounted for as it is revealed through a receding-horizon framework.
Fourth, we assume that each vehicle carries only a single customer. This mode of operation is consistent with trends in current mobility-on-demand systems, such as taxis, Uber, and Lyft. The extension to ride-sharing in AMoD systems is an interesting direction for future research.
Finally, we assume for the sake of simplicity that all customers have similar preferences in terms of travel comfort and value of time. However, the model proposed in this paper can readily be extended to capture multiple classes of customers, each characterized by a different preference profile and modeled by a distinct network flow.
%% !TEX root = ../Salazar.Rossi.ea.ITSC18.tex
\section{A Pricing and Tolling Scheme for \gls{abk:iamod}}\label{sec:pricing}
The mesoscopic modeling approach presented in Section~\ref{subsec:model} along with the objective function presented in Section~\ref{subsec:objective} assume an idealized scenario whereby the objectives of all stakeholders are aligned with the objective of maximizing social welfare.
In reality, customers decide on their routes so as to maximize their private welfare, i.e., minimizing their travel time and the price paid for their trip, and AMoD fleet operators control their fleet's operations so as to maximize their profits.

In this section, we propose a pricing and tolling scheme to align the goals of selfish agents with the objective of maximizing social welfare, as defined in Section~\ref{subsec:objective}. Section~\ref{subsec:agents} introduces the self-interested agents participating in the \gls{abk:iamod} market, while Section~\ref{subsec:pricingscheme} details our pricing and tolling scheme and Section~\ref{subsec:genequi} proves that the social optimum coincides with an equilibrium induced by the proposed scheme.

\subsection{Self-interested Agents}\label{subsec:agents}
%Formally,
We model the \gls{abk:iamod} market as a perfect market with three classes of participants: The municipal transportation authority, \gls{abk:iamod} customers and \gls{abk:amod} operators. Our key assumption of a perfect \gls{abk:iamod} market entails that no individual customer or \gls{abk:amod} operator is able to unilaterally influence the transportation prices, which are rather set by the market equilibrium~\cite{Mas-ColellWhinstonEtAl1995}.

\textbf{The municipal transportation authority} monitors the operations of the I-AMoD system and sets (i) fares in the subway system and (ii) road usage tolls in the road network with the goal of maximizing social welfare.
Tolls in the road network may be interpreted as congestion surcharges, whereas prices in the public transportation network account for the amortized operational cost of the transportation system.
Specifically, the  transportation authority sets a fare $\pSubway$ for each arc $\arc\in\setOfArcsSubway$ in the public transportation network and a toll $ \tRoad$ for each arc $\arc\in \setOfArcsRoad$ in the road network.

\textbf{I-AMoD customers} travel within the network.
Each customer is associated with a request $r_m$ and selects an intermodal route from her origin to her destination.
In our mesoscopic perspective, route selection entails choosing a commodity flow $f_m(\cdot,\cdot)$ so as to satisfy Eq.~\eqref{eq:customerbalance}.
If the customer decides to travel on public transit, she pays the price $ \pSubway$ set by the municipal transportation authority for each arc $\arc\in\setOfArcsSubway$ traversed. If the customer chooses to use the AMoD system for all or parts of her trip, she pays a fare to the AMoD operator. The fare is composed of (i) a charge $ \pOrigin{i}$ associated with the selected pick-up location, (ii) a charge $\pDest{j}$ associated with the destination location, and (iii) a charge  $ \pRoad$ for each road arc $\arc\in \setOfArcsRoad$  traversed. Given the mesoscopic perspective of our study, we neglect common user-centric modeling approaches that account for individual cost functions. 
In line with current practice~\cite{Google2018}, we assume that customers select their routes by using navigation apps, which compute routes by accounting for an aggregate model of the customers' route preferences.
Specifically, we set the customer's objective as the maximization of her welfare, defined as the sum of (i) the travel time multiplied by the value of time $\vTime$ and (ii) the cost of her trip as the cumulative sum of the fares paid along the route. For consistency with the definition of social cost as given in Section \ref{subsec:objective}, and to enable the design of a socially-optimal pricing and tolling scheme, we include the same quadratic regularization term as in Eq.~\eqref{eq:costM}.
Formally, \gls{abk:iamod} customers solve the problem

%\begin{equation}
\begin{align}
\label{eq:customerproblem}
\min_{\flow{\cdot}{\cdot}}  & \vTime\cdot \sum_{\arc\in \setOfArcs } t_{ij}\cdot \flow{i}{j} + \vQuadratic\cdot \sum_{\arc\in\setOfArcs}\flow{j}{j}^2  \nonumber \\
& +  \sum_{j\in\setOfVerticesRoad}  \pOrigin{j} \cdot\left( \sum_{k: (j,k)\in\setOfArcsRoad} \flow{j}{k} - \sum_{i: \arc\in\setOfArcsRoad} \flow{i}{j}\right)^+  \nonumber \\
& + \sum_{j\in\setOfVerticesRoad}  \pDest{j} \cdot \left( \sum_{i: \arc\in\setOfArcsRoad} \flow{i}{j} - \sum_{k: (j,k)\in\setOfArcsRoad} \flow{j}{k}\right)^+ \nonumber \\
&+ \sum_{\arc\in \setOfArcsRoad }  \pRoad\cdot \flow{i}{j} + \sum_{\arc\in \setOfArcsSubway }  \pSubway \cdot\flow{i}{j} \nonumber \\
%\hfill  \nonumber \\
\vspace{4mm}
\text{s.t. } & \text{Eq.~\eqref{eq:customerbalance}},
\end{align}
%\end{equation}
where $(\cdot)^+=\max(\cdot,0)$. The first term in the cost function corresponds to the customer's value of time, the third and fourth term capture the origin and destination charges in the AMoD network, the fifth term denotes the arc-based charge in the AMoD network, and the last term is the fare paid to the subway network.

\textbf{AMoD operators} service customer requests and collect fares from the customers. The operators also control the rebalancing vehicles' routes to ensure that vehicles are available to service customer requests. Without loss of generality, we can combine the rebalancing flows for each AMoD operator into an individual rebalancing flow. Hence, for the sake of simplicity, for the rest of the paper we will fold the \gls{abk:amod} operators into a unique operator. A detailed explanation is to be found in Section~\ref{subsec:pricing:discussion}.

Tolls $\tRoad$ set by the municipal transportation authority are levied for vehicles for each road arc traversed.
Consistent with the assumption of a perfect market, the AMoD operator is unable to influence the AMoD prices $\pOrigin{i}, \pDest{j}$, and $\pRoad$;
rather, these prices are determined by the market equilibrium. Accordingly, the AMoD operator's goal of maximizing revenue is equivalent to the goal of minimizing operating expenses. Again, we include the same quadratic regularization term in Eq.~\eqref{eq:costM} within the AMoD operator's cost to ensure strict convexity. Therefore, the  \gls{abk:amod} operator solves the problem

\begin{small}
\vspace{-1mm}
	%\begin{equation}
	\begin{align}
	\label{eq:operatorproblem}
	\min_{\flowReba{\cdot}{\cdot}} & \sum_{\arc \in \setOfArcsRoad} \left(\vDistR\cdot d_{ij} +\vEnergy\cdot\energyR+  \tRoad\right)\cdot \flowReba{i}{j}+ \vQuadratic\cdot  \flowReba{i}{j}^2\nonumber \\
	\text{s.t. } & \text{Eq.~\eqref{eq:flowreba}}.
	\end{align}
%	\end{equation}
\end{small}
\vspace{-1mm}

\subsection{A Pricing and Tolling Scheme}\label{subsec:pricingscheme}
\label{subsec:pricing}
The subway fares $\pSubway$ and the road tolls $ \tRoad$ can be considered as steering variables that a public stakeholder can optimize in order to align the goals of self-interested customers and the AMoD operator with the  objective defined in Section \ref{subsec:objective}.

We denote the dual variables associated with the constraints in Problem \eqref{eq:IAMoD} as $\dualCust$ for customers balance 
 \eqref{eq:customerbalance}, $\dualVeh$ for vehicles balance \eqref{eq:flowreba},
$\dualcR$ for road arcs capacity \eqref{eq:caproads}, and $\dualcS$ for subway arcs capacity \eqref{eq:cappublic}.
We propose the following pricing and tolling scheme.
The subway fares $ \pSubway$ are set as the sum of the public transit operational cost $\vDistS\cdot d_{ij}$ and the dual variables of the subway congestion constraint $\dualcS$:
\begin{equation}
\pSubway=\vDistS\cdot d_{ij} +\dualcS(i,j).   \label{eq:prices:subwaytolls}
\end{equation}
The road tolls $ \tRoad$ are set as the dual variables of the road congestion constraint $\dualcR$:
\begin{equation}
\tRoad= \dualcR(i,j) .\label{eq:prices:roadtoll}
\end{equation}

\subsection{A General Equilibrium}\label{subsec:genequi}
The following theorem shows that the pricing and tolling scheme proposed in Section~\ref{subsec:pricing} ensures that
an optimal solution to the \gls{abk:iamod} Problem~\eqref{eq:IAMoD} coincides with a general equilibrium for the perfect market described in Section~\ref{subsec:agents}.
\begin{theorem}[Optimal Pricing and Tolling Scheme]
\label{thm:general-equilibrium}
Consider the following prices.
The destination charges $ \pDest{i}$ are equal to the dual variables of the vehicle conservation constraint $\dualVeh(i)$, while the origin charges $ \pOrigin{i}$ are set as the opposite, that is,
\begin{equation}
\pOrigin{i} = -\pDest{i} = -\dualVeh(i). \label{eq:prices:AMoDorigin}
\end{equation}
The road arc charges $ \pRoad$ are equal to the sum of the vehicles' operating costs $\vDistR\cdot d_{ij}+\vEnergy\cdot\energyR$ and the road tolls $ \tRoad$, that is,
\begin{equation}
\pRoad= \vDistR\cdot d_{ij} +\vEnergy\cdot\energyR+   \tRoad. \label{eq:prices:AMoDroad}
\end{equation}
Consider an optimal solution $\big\{\{\flow{\cdot}{\cdot}\}_m,\flowReba{\cdot}{\cdot}\big\}$ to the I-AMoD problem. Also, consider a perfect market where self-interested customers plan their routes by solving Problem \eqref{eq:customerproblem}, a self-interested AMoD operator plans routes for the rebalancing vehicles by solving Problem \eqref{eq:operatorproblem}, and a municipal transportation authority sets public transit prices and road tolls  according  to \eqref{eq:prices:subwaytolls}-\eqref{eq:prices:roadtoll}.
Then the AMoD prices \eqref{eq:prices:AMoDorigin}-\eqref{eq:prices:AMoDroad} are market-clearing prices and the optimal solution $\big\{\{\flow{\cdot}{\cdot}\}_m,\flowReba{\cdot}{\cdot}\big\}$ is a general equilibrium for the \gls{abk:iamod} market.
\end{theorem}
\emph{Proof sketch:}
The proof relies on showing that satisfaction of the KKT conditions for the I-AMoD Problem~\eqref{eq:IAMoD} implies satisfaction of the KKT conditions for the customers' optimal routing Problem~\eqref{eq:customerproblem} and the KKT conditions for the AMoD operator's optimal rebalancing Problem~\eqref{eq:operatorproblem}. A rigorous proof is reported 
\ifextendedversion
in Appendix~\ref{sec:appendix:proof}.
\else
in the extended version of this paper \cite{SalazarRossiEtAl2018}.
\fi

\subsection{Discussion}
\label{subsec:pricing:discussion}
A few comments are in order.
First, in the setting of a general equilibrium, we assume that the AMoD operators have no pricing power. That is, no individual AMoD operator is able to single-handedly influence the fares paid by the AMoD customers.
This assumption holds true in the case where multiple operators of similar size compete for customers' transportation demands, and it is arguably realistic in several urban environments. For reference, no fewer than five app-based mobility-on-demand operators (Uber, Lyft, Juno, Curb, and Arro) offer mobility-on-demand services in Manhattan at the present day.
Second, in this paper, the operations of all AMoD operators are captured through a single rebalancing flow and a single set of customer-carrying flows on road arcs for simplicity and ease of notation. However, the model does \emph{not} assume that a single AMoD operator is present. Indeed, a treatment where different operators control different subsets of vehicles, each associated with a rebalancing flow, would result in the same equilibrium.
To see this, note that customers, not AMoD operators, choose which operator to use by selecting the customer-carrying flows $\{\flow{\cdot}{\cdot}\}_m$, and the operators do not compete on prices. For a given set of customer requests and given road tolls, each AMoD operator solves a smaller version of Problem~\eqref{eq:operatorproblem}. Critically, the optimization problems of different AMoD operators are not coupled. That is, the decision variables of each operator do not affect the other operators' problems. Thus, solving Problem~\eqref{eq:operatorproblem} is equivalent to solving each AMoD operator's optimization problem, and the result in Theorem~\ref{thm:general-equilibrium} holds true for each individual AMoD operator.
Third, we assume that the routes followed by customer-carrying AMoD vehicles are set by the customers themselves through the navigation apps.
In practical implementations, the customers may be able to choose only among a limited set of possible routes, for example between a direct route that incurs congestion tolls and a longer, less congested and thus cheaper route.
Such more sophisticated route selection models are left for future research.
Finally, we use the cost function~\eqref{eq:customerproblem} to model customers' behavior.
Although such an approach does not entail the level of detail of a user-centric approach~\cite[Ch.~4]{GentileNoekel2016}, it suffices for the mesoscopic perspective of this study.
%% !TEX root = ../paper.tex
\section{Results}\label{sec:results}
In this section, we assess the performance achievable by the \gls{abk:iamod} system in terms of travel time, costs, and emissions in a real-world case study for Manhattan. Section~\ref{sec:results:scenario} details this case study. Thereafter, we present the optimal solution for the \gls{abk:iamod} system in Section~\ref{sec:results:iamod} and compare it to the optimal solution for the \gls{abk:amod} system operating in isolation in Section~\ref{sec:results:compare}.

\subsection{Case Study}\label{sec:results:scenario}
\begin{figure}[t]
	\centering\includegraphics[width=\columnwidth]{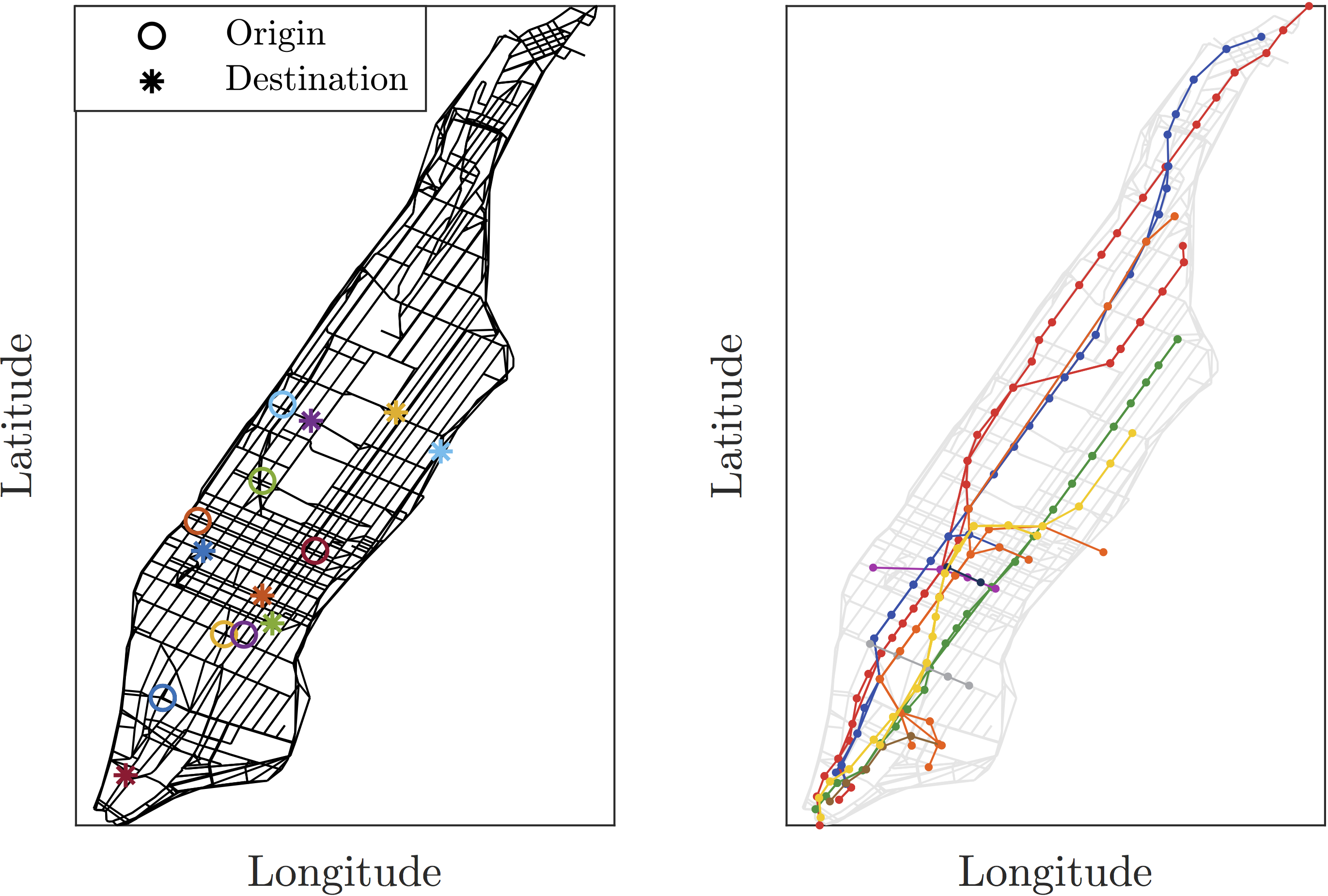}
	\caption{Road map and subway lines of Manhattan with exemplary origin-destination pairs.}
	\label{fig:RoadMap}
\end{figure}

We focus on the Manhattan transportation network in New York City  shown in Fig.~\ref{fig:RoadMap}.
The transportation requests considered are the actual 53'932 taxi rides which took place on March 1, 2012 between 6 and 8 p.m.\ in Manhattan (courtesy of the New York Taxi and Limousine Commission). The 6'772 origin-destination pairs of the trips are placed on the pedestrian digraph. We derive the road network from OpenStreetMap data\cite{HaklayWeber2008}, and define the capacity of each street to be proportional to the number of lanes multiplied by the road's speed limit. Since taxis constitute only a fraction of the cars in Manhattan, we set the road capacity available to the AMoD fleet to be a fraction of the overall road capacity. Without restricting ourselves to a fixed number, we perform a parametric study where the scaling factor of the road capacity is varied between 0 and 10\%. The topology of the walking network is similar to the topology of the road graph with the key difference that pedestrians, unlike vehicles, can travel along both directions on every road link. We connect every node in the road digraph to its equivalent in the walking network, modeling the customers' ability to hail an AMoD ride. We assume the subway network to be the only public transportation system, in line with the fact that the subway network is the dominant public transit mode in Manhattan. This way, we provide a first order assessment of \gls{abk:iamod}. We construct the public transportation digraph using the geographical location of the lines and the stops found in the NYC Open Data database~\cite{OpenDataNYCSubway2018} as well as the time schedules of the MTA~\cite{MTA2018}. We connect every node in the public transportation network to the geographically closest node of the walking digraph. 
We set the time to transfer from a road node or a subway stop to a walking node, which models the time required to exit an AMoD vehicle or a subway station, to one minute. We assume that two minutes are required to go from a pedestrian to a road node and get into an \gls{abk:amod} vehicle, which is in line with the average time to hail a ride in Manhattan~\cite{MosendzSender2014}. The time to transfer from a walking node to a subway line equals one minute plus one half of the periodicity of the line. 

Table~\ref{tab:data} summarizes the remaining parameters used in our case study: According to the DOT guidelines, we set the value of time to \unit[24.40]{USD/h}~\cite{Endorf2015} and the operational cost for the \gls{abk:amod} vehicles excluding electricity costs to \unit[0.486]{USD/mile}~\cite{BTS2016}. For the subway, the operational cost per passenger mile is equal to \unit[0.47]{USD/mile}~\cite{NeffDickens2017}. We assume the \gls{abk:amod} fleet to be composed of electric lightweight vehicles and derive their parameters from~\cite[Ch.~2]{GuzzellaSciarretta2007}. The cost of electricity is set to~\unit[0.247]{USD/kWh}~\cite{Rueb2017}. We directly relate the energy consumption to the CO\textsubscript{2} emissions based on the current electricity sources of the state of New York~\cite{Watttime2018}.
\begin{table}[t]
	\caption{Numerical data for the case study}
	\begin{center}
		\begin{tabular}{llll}\label{tab:data}
			%\midrule
			Parameter & Variable & Value & Source\\
			\midrule
			Value of time & $\vTime$  & \unit[24.40]{USD/h} & \cite{Endorf2015} \\
			Vehicle operational cost & $\vDistR$ & \unit[0.486]{USD/mile} & \cite{BTS2016}\\
			Subway operational cost & $\vDistS$ & \unit[0.47]{USD/mile} & \cite{NeffDickens2017}\\
			Cost of electricity & $\vEnergy$ & \unit[0.247]{USD/kWh}  & \cite{Rueb2017}\\
			Air density & $\rho_\mathrm{air}$ & \unit[1.25]{kg/m$^3$} & \cite[Ch.~2]{GuzzellaSciarretta2007}\\
			Frontal drag coefficient & $c_\mathrm{d}\cdot A_\mathrm{f}$ & \unit[0.4]{m$^2$} & \cite[Ch.~2]{GuzzellaSciarretta2007}\\
			Rolling friction coefficient & $c_\mathrm{r}$ &0.008 & \cite[Ch.~2]{GuzzellaSciarretta2007}\\
			Mass of the vehicle & $m_\mathrm{v}$ & \unit[750]{kg} & \cite[Ch.~2]{GuzzellaSciarretta2007}\\
			Tank-to-wheel efficiency& $\eta_\mathrm{EV}$ & \unit[72]{\%} & \cite[Ch.~2]{GuzzellaSciarretta2007}\\
			\bottomrule
		\end{tabular}
	\end{center}
\end{table}

For each of the scenarios presented in the next Sections~\ref{sec:results:iamod} and~\ref{sec:results:compare}, the quadratic optimization Problem~\eqref{eq:IAMoD} was solved on commodity hardware (Intel Core i7, \unit[16]{GB} RAM) using Gurobi 7.5.2 in less than 10 minutes.

\subsection{Optimal Solution for the \gls{abk:iamod} System}\label{sec:results:iamod}
\begin{figure}[t]
	\centering\includegraphics[width=\columnwidth]{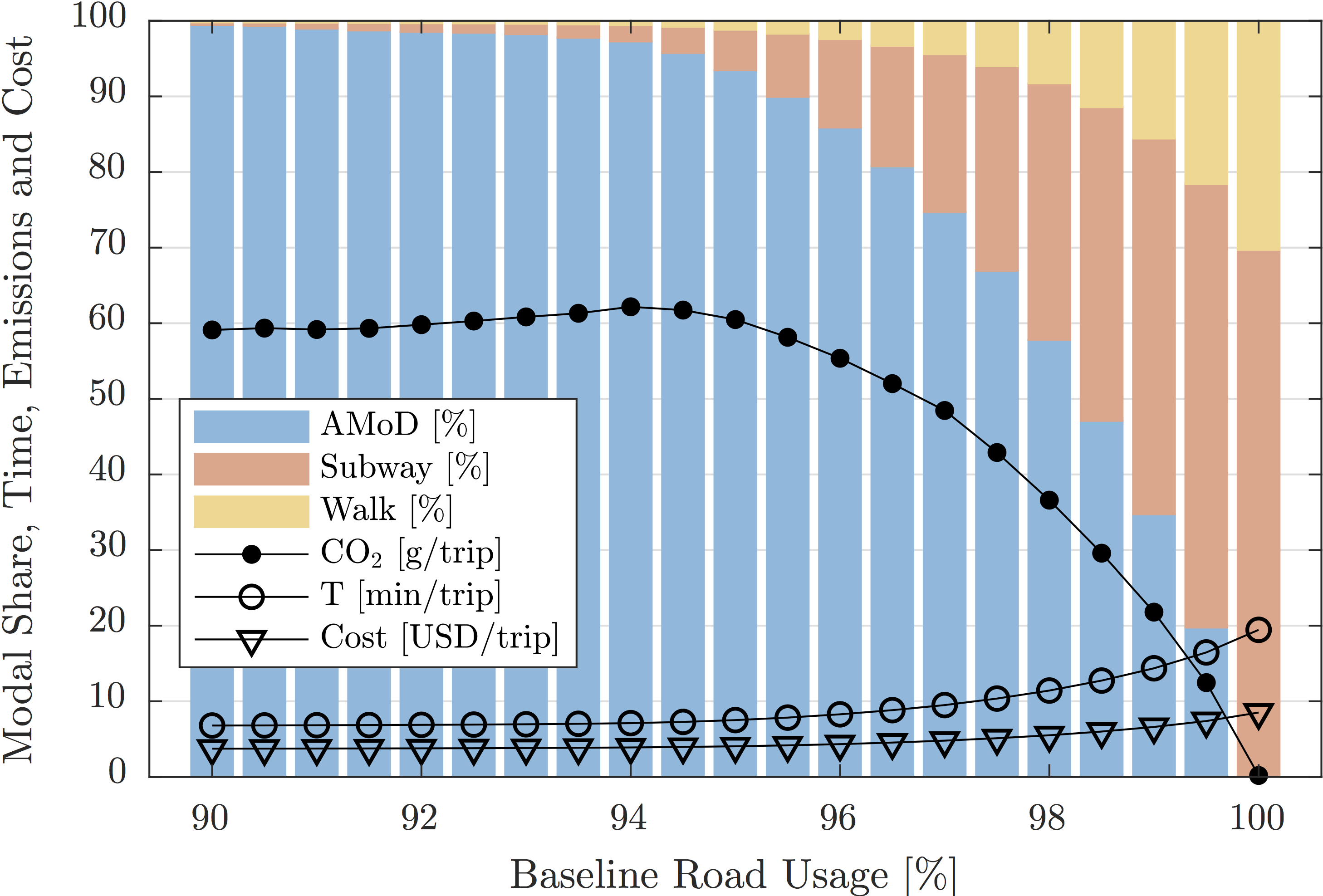}
	\caption{Distance-based modal share for the \gls{abk:iamod} system.}
	\label{fig:dshare_iamod}
\end{figure}

In this section, we study the performance achievable by the \gls{abk:iamod} system by solving Problem~\eqref{eq:IAMoD} for different levels of road usage.
Specifically, a 100\% baseline road usage sets the road capacity to zero, whereas a 90\% usage means that 10\% of the empty road capacity is available to AMoD vehicles. 
Fig.~\ref{fig:dshare_iamod} shows the distance-based modal share for different levels of road usage, together with the average travel time, CO\textsubscript{2} emissions, and monetary cost $\costM$. As the available road capacity lowers, the subway utilization increases. With a further decreasing road capacity the walking distance also increases because the subway cannot fully replace a point-to-point means of transportation. Nevertheless, the travel time and monetary costs remain well contained, while emissions drop significantly.
\begin{figure}[t]
\centering\includegraphics[width=\columnwidth]{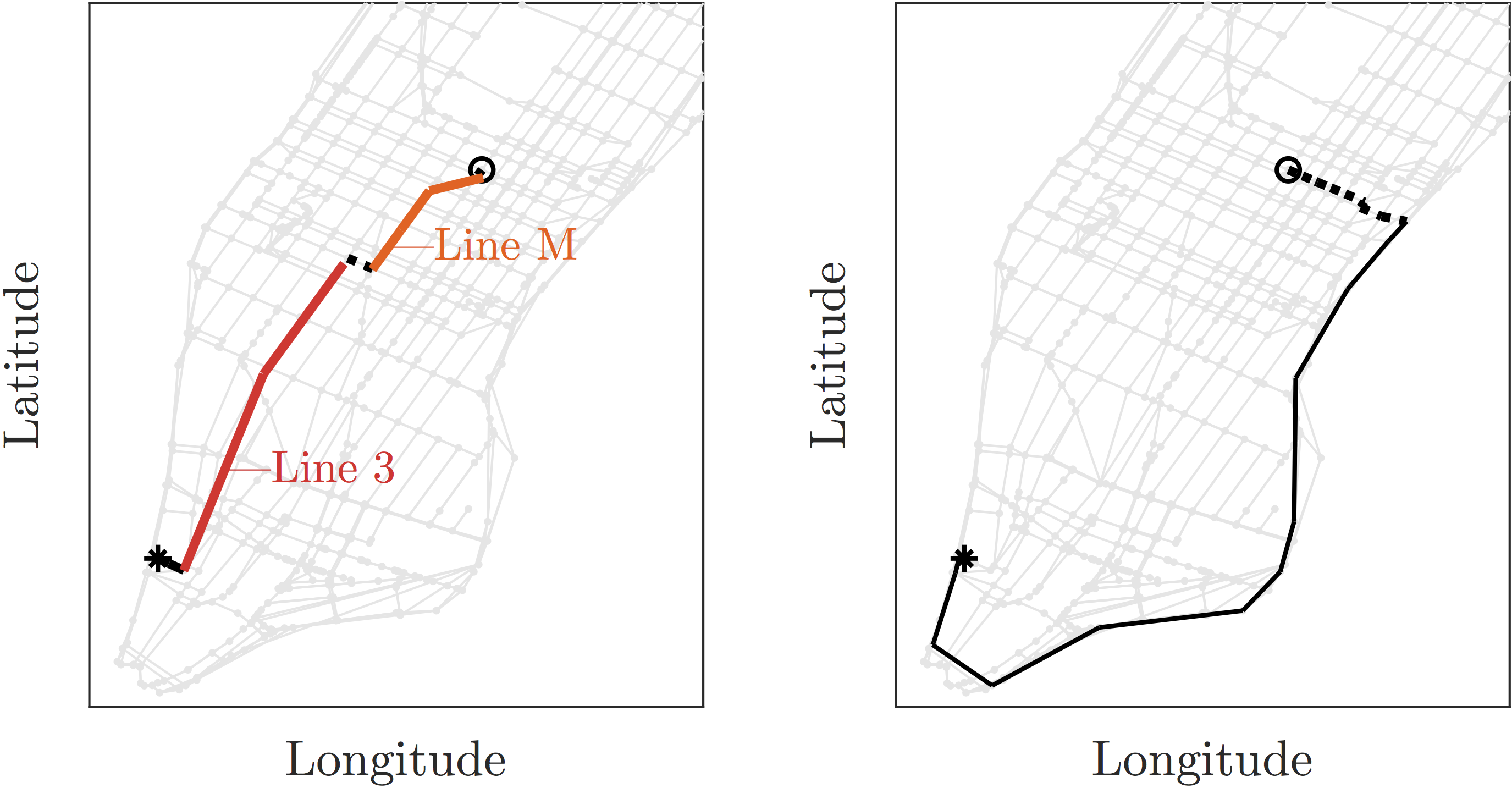}
\caption{Optimal paths for \gls{abk:iamod} (left) and \gls{abk:amod} (right). Segments traveled by car are solid black, by foot dashed black and by subway solid colored as the corresponding line.}
\label{fig:travel_iamod}
\end{figure}
The left part of Fig.~\ref{fig:travel_iamod} shows an exemplary optimal path under a 98\% baseline road usage, where different means of transportation are used. The customer flow travels with the M~line from 5 Avenue / 53 Street to 34 Street - Herald Square, walks one block, and takes the 3 line from 34 Street - Penn Station to Chambers Street, reaching her destination by \gls{abk:amod} car.

\begin{figure}[t]
	\centering
	\begin{minipage}{0.49\columnwidth}
		\includegraphics[width=\columnwidth]{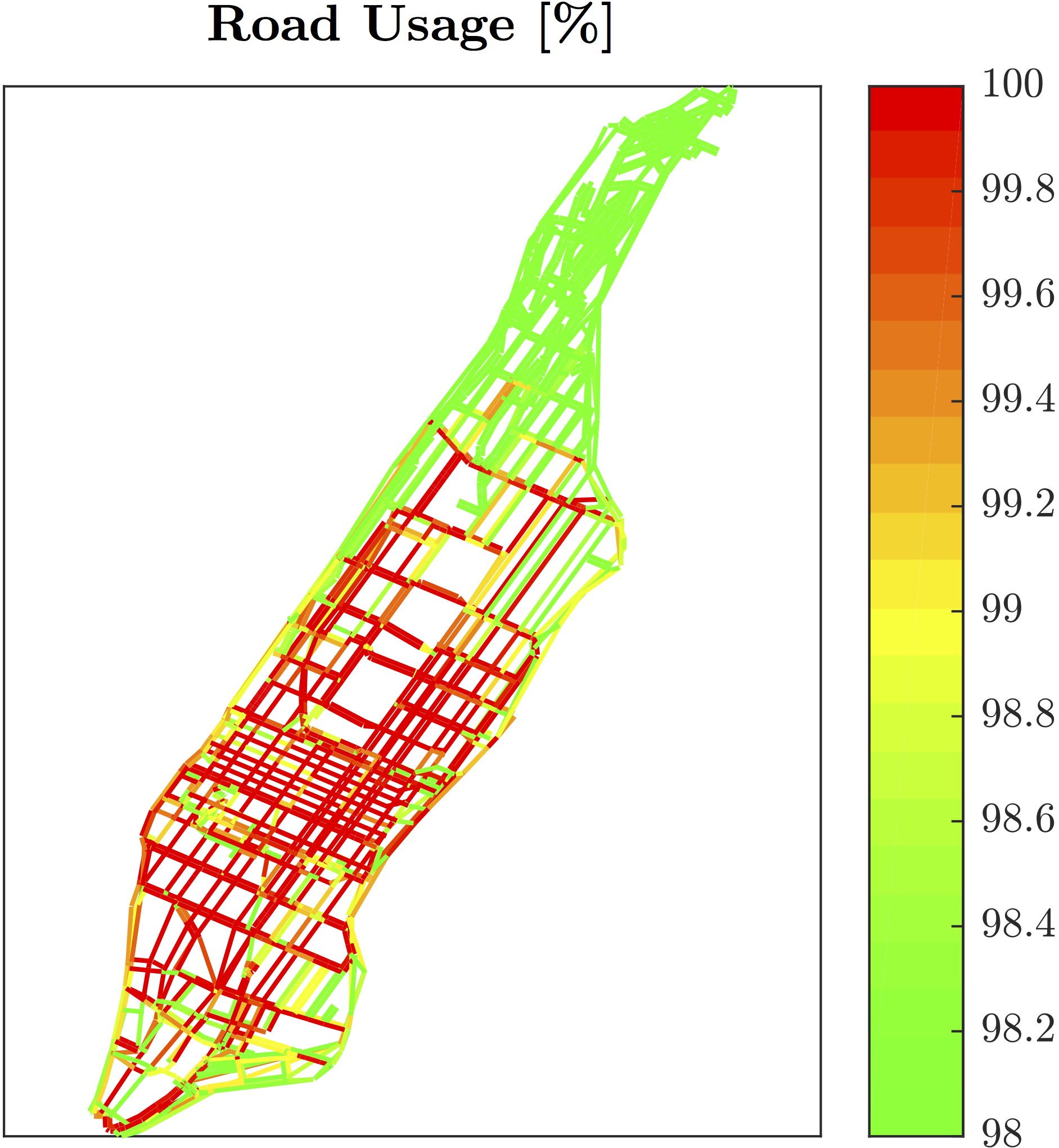}
	\end{minipage}
	\begin{minipage}{0.49\columnwidth}
		\includegraphics[width=\columnwidth]{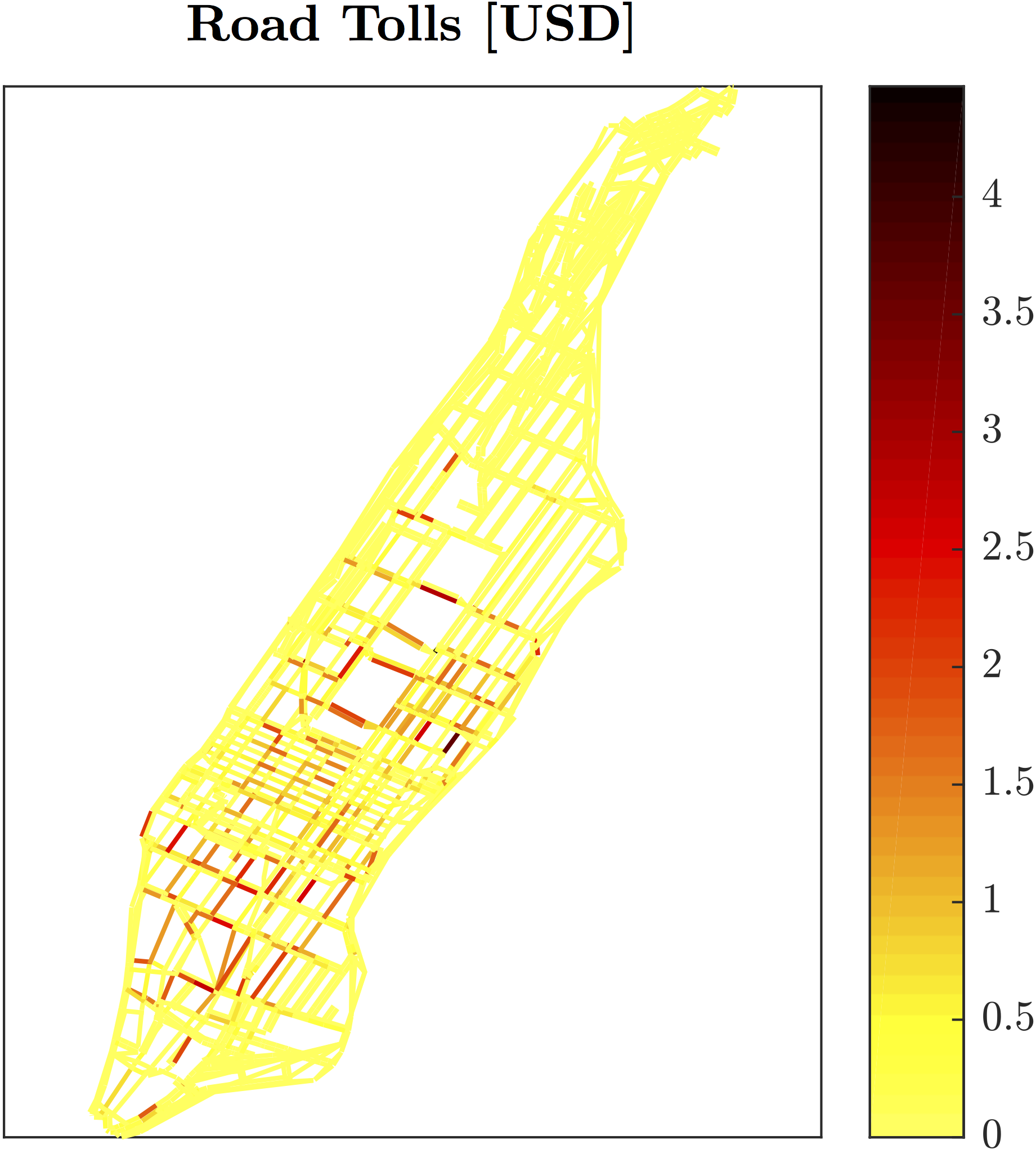} 
	\end{minipage}
	\caption{Road usage and optimal tolls for the \gls{abk:iamod} system.}
	\label{fig:I-AMoD98}
\end{figure}

Fig.~\ref{fig:I-AMoD98} shows the road usage and the road tolls $\tRoad$ for the same baseline scenario.
The average cost of fares paid due to tolls along trips corresponds to almost 2 USD, which is interestingly in line with current proposals to tax mobility-on-demand vehicles with a congestion surcharge of 2 to 5 USD per trip~\cite{HuWang2018}.

\subsection{Comparison with the \gls{abk:amod} System}\label{sec:results:compare}
\begin{figure}[t]
	\centering\includegraphics[width=\columnwidth]{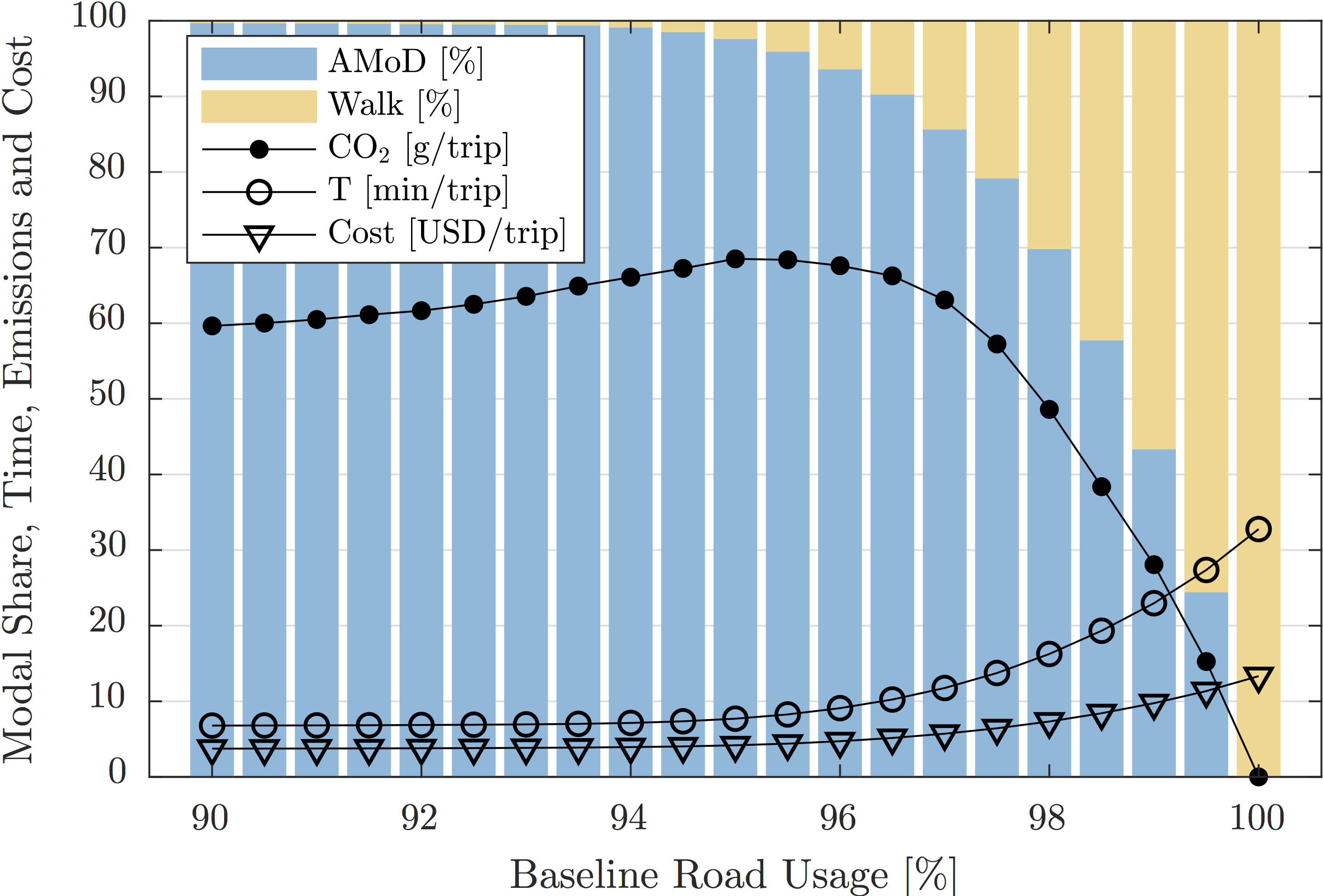}
	\caption{Distance-based modal share for the \gls{abk:amod} system.}
	\label{fig:dshare_amod}
\end{figure}

\begin{figure}[t]
\centering
\begin{minipage}{0.49\columnwidth}
	\includegraphics[width=\columnwidth]{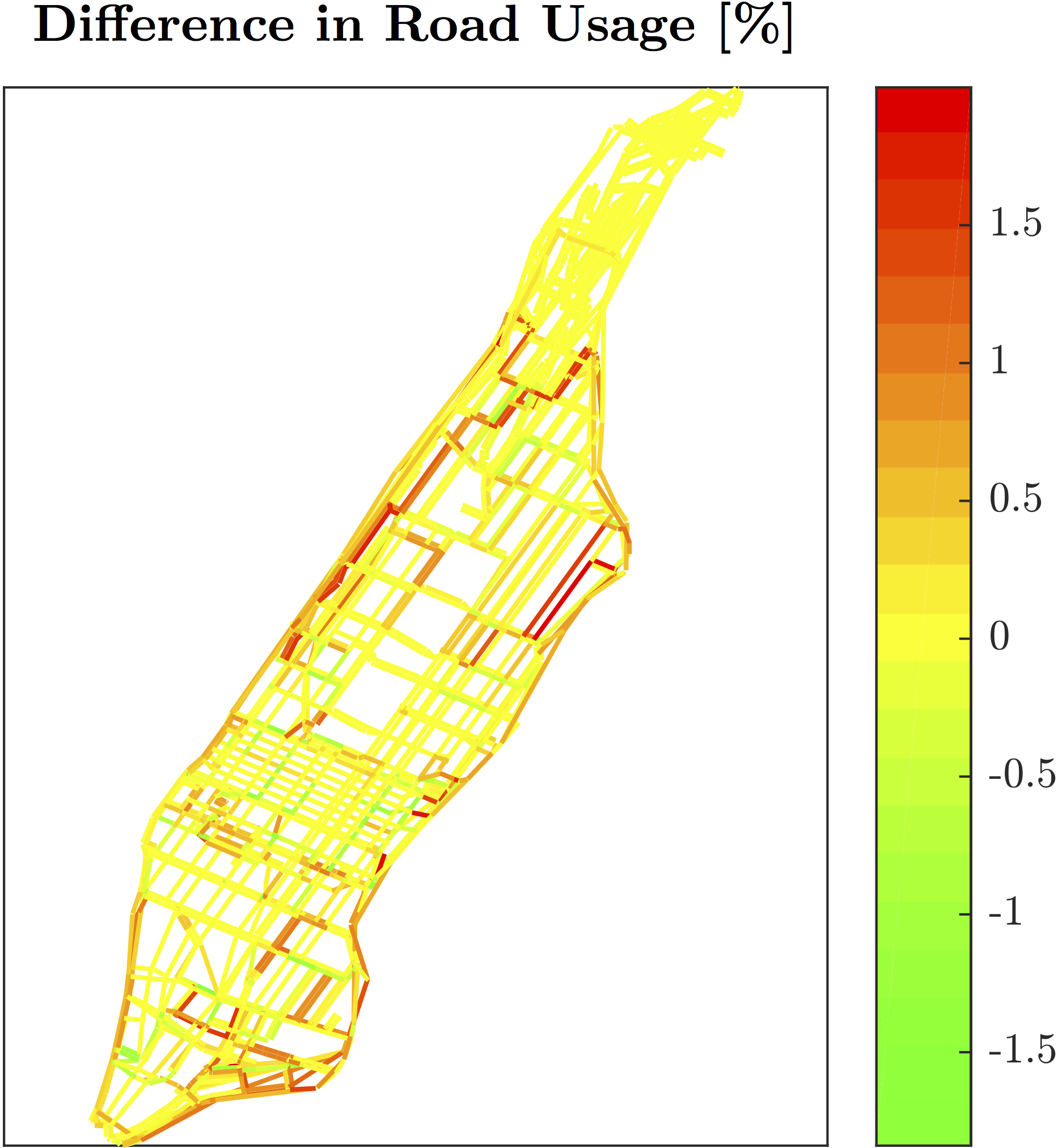}
\end{minipage}
\begin{minipage}{0.49\columnwidth}
	\includegraphics[width=\columnwidth]{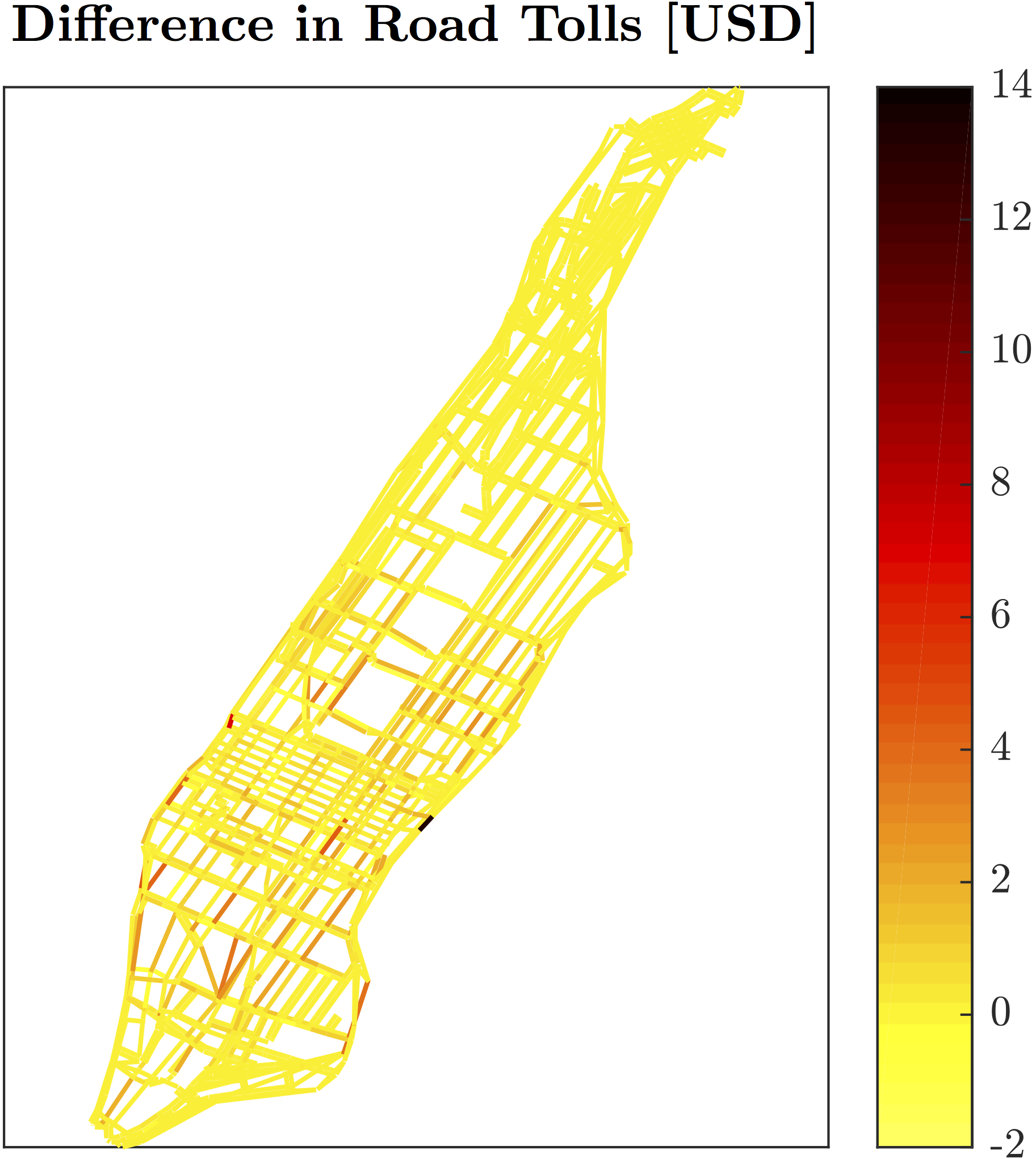}
\end{minipage}
\caption{Difference in road usage and optimal tolls.}
\label{fig:amodvsiamod_C98}
\end{figure}

We analyze the achievable performance of the \gls{abk:amod} system operating in isolation for the same levels of traffic as in the previous Section~\ref{sec:results:iamod} and compare it to that of the \gls{abk:iamod} system. Within our framework, the \gls{abk:amod} solution is obtained by simply setting the subway capacity of the \gls{abk:iamod} system to zero. The ``infinite" capacity of the pedestrian network guarantees feasibility even under extremely congested conditions. 
Fig.~\ref{fig:dshare_amod} shows the modal share in this scenario.
As the baseline road usage increases and the available road capacity decreases, the average travel time increases due to longer routes and larger walking distances. The monetary cost follows a similar trend, whereas emissions reach a maximum due to longer customer and rebalancing trips before falling to zero when the roads are fully saturated and all trips are traveled on foot.
The right side of Fig.~\ref{fig:travel_iamod} shows an exemplary optimal path with the same origin-destination pair as the path of the \gls{abk:iamod} system shown on the left side.
The customer flow walks until reaching a vehicle which drives her through the FDR to her destination.
\begin{figure}[t]
	\centering\includegraphics[width=\columnwidth]{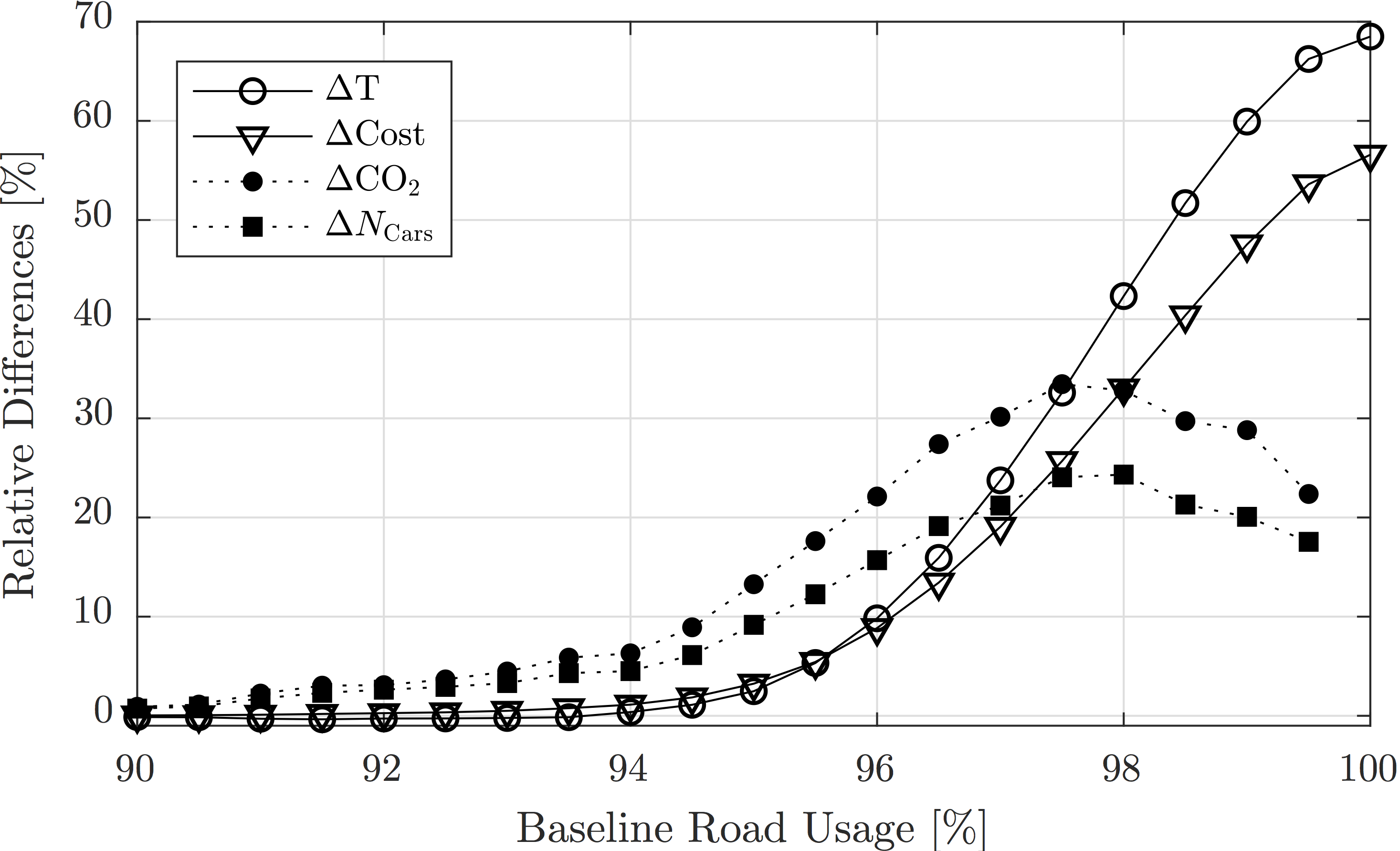}
	\caption{Relative difference in travel time, cost, emissions. and number of cars between \gls{abk:amod} and \gls{abk:iamod}.}
	\label{fig:amodvsiamod_TE}
\end{figure}
Fig.~\ref{fig:amodvsiamod_TE} shows the relative decrease in average travel time, CO\textsubscript{2} emissions, and monetary cost achievable by coordinating the \gls{abk:amod} fleet with the public transportation network. As the road availability decreases, the difference in travel time and cost increases by more than 40\%, while the emissions get reduced by almost 30\%. Fig.~\ref{fig:amodvsiamod_C98} shows that due to the severe congestion constraints, the difference in road usage is minor. However, the higher road tolls would cause an average surcharge of almost 6 USD per trip, 200\% more than for the \gls{abk:iamod} system.
Overall, \gls{abk:iamod} results in shorter travel times, fewer vehicles, and much lower emissions and tolls.
%% !TEX root = ../Salazar.Rossi.ea.ITSC18.tex
\section{Conclusion}\label{sec:conclusion}
In this paper, we explored the possibility of coordinating different modes of transportation in congested urban environments in order to satisfy travel requests whilst maximizing social welfare. We presented a multi-commodity network flow model for an autonomous mobility-on-demand system that cooperates with the public transportation network.
Besides computing optimal customer and vehicle routes, we designed a pricing and tolling scheme that was proven able to steer selfish agents to the social optimum under the assumption of a perfect market.
We presented a case study for Manhattan and we showed that optimizing the customer and the rebalancing routes by jointly accounting for the \gls{abk:amod} fleet and the public transportation system can dramatically reduce travel time, costs, and emissions. The optimal road tolls computed within our model are quantitatively in line with the surcharges for ride-hailing trips recently discussed by the New York City municipal authority and significantly lower than for the \gls{abk:amod} system in isolation.

This work opens the field for several research directions.
First, we would like to design an operational algorithm to compute the optimal customer and rebalancing routes at the microscopic level in real-time.
Second, it is of interest to extend our model to capture stochastic effects such as time-varying congestion, public transportation delays, and variable customer demand.
Third, we would like to combine our model with a power-in-the-loop \gls{abk:amod} model~ \cite{RossiIglesiasEtAl2018}, in order to investigate to which extent intermodality can improve the interaction with the electric power grid.
Fourth, we would like to explore more human-centered optimization objectives such as travel comfort and switch-over costs.
Finally, we plan to study the impact of \gls{abk:iamod} systems on different cities with diversely advanced public transportation networks.
\section*{Acknowledgments}
We would like to thank Dr.\ Daniele Vigo and Dr.\ Guido Gentile for the fruitful discussions,
and Dr.\ Ilse New for her assistance with the proofreading and useful advice.
The first author would like to thank Dr.\ Lino Guzzella for his support.
This research was supported by the National Science Foundation under CAREER Award CMMI-1454737 and the Toyota Research Institute (TRI). This article solely reflects the opinions and conclusions of its authors and not NSF, TRI, or any other entity.
This paper is dedicated to Stella.
%%%%%%%%%%%%%%%%%%%%%%%%%%%%%%%%%%%%%%%%%%%%%%%%%%%%%%%%%%%%%%%%%%%%%%%%%%%%%%%%
\bibliographystyle{IEEEtran}
\bibliography{../../../bib/main,../../../bib/ASL_papers}
%\bibliography{paper.bbl}
\ifextendedversion
%% !TEX root = ../Salazar.Rossi.ea.ITSC18.tex
\clearpage
\begin{appendix}
\subsection{Proof of Theorem 3.1}\label{sec:appendix:proof}
\begin{proof}
	The proof relies on showing that the prices and tolls \eqref{eq:prices:subwaytolls}-\eqref{eq:prices:AMoDroad} align the incentives of self-interested agents with the social optimum. Namely, the origin and destination charges ensure that customers are charged for the amount of imbalance (and therefore for the additional rebalancing trips) that their trip induces in the \gls{abk:amod} system; the subway prices and road prices ensure that customers account for induced congestion and overcrowding, and operational costs in their decision-making process; and the road tolls ensure that a selfish \gls{abk:amod} system operator takes into account the congestion caused by its vehicles.
	
	Specifically, we show that the optimal solution to the \gls{abk:iamod} Problem~\eqref{eq:IAMoD} is also 
	the
	%\msmargin{the}{\\Right?}
	%	\frmargin{}{I am not convinced. We show that, for given prices, the AMoD operator and the customers have an unique optimum. But in principle there could be other sets of prices for which the AMoD operator and the customers are in equilibrium. Right? I may be confused, of course. But now we prove that this is the only equilibrium \emph{for these prices}, I think.}
	general equilibrium for customers and the \gls{abk:amod} system operator if prices and tolls are computed according to \eqref{eq:prices:subwaytolls}-\eqref{eq:prices:AMoDroad}; specifically, if $\big\{\{\flow{\cdot}{\cdot}\}_m,\flowReba{\cdot}{\cdot}\big\}$ is an optimal solution to \eqref{eq:IAMoD} with associated prices \eqref{eq:prices:subwaytolls}-\eqref{eq:prices:AMoDroad}, then $\{\flow{\cdot}{\cdot}\}_m$ is an optimal solution to \eqref{eq:customerproblem} for each $m\in\setOfRequestsNumber$, and $\flowReba{\cdot}{\cdot}$ is an optimal solution to \eqref{eq:operatorproblem}. Since all three problems are strictly convex, they are guaranteed to have a unique globally optimal solution. Therefore, it is sufficient to show that satisfaction of the KKT coditions for the \gls{abk:iamod} Problem~\eqref{eq:IAMoD} implies satisfaction of the KKT conditions for the customers' optimal routing Problem~\eqref{eq:customerproblem} and the \gls{abk:amod} operator's optimal rebalancing problem~\eqref{eq:operatorproblem}~\cite{BoydVandenberghe2004}.

	The KKT stationarity conditions for the \gls{abk:iamod} Problem~\eqref{eq:IAMoD} are:
%	\begingroup
%	\allowdisplaybreaks

	\begin{small}
	\begin{subequations}
		\begin{align}
		&\left(\vTime\cdot \traveltime_{ij} + \vDistR\cdot  d_{ij}+\vEnergy\cdot\energyR\right) + \dualCust(j)
		- \dualCust(i) +\dualVeh(j) \span \nonumber\\
		& - \dualVeh(i) + \dualcR(i,j)+2\cdot\vQuadratic\cdot\flow{i}{j}^\star =0  &\forall m\in \setOfRequestsNumber, \arc\in\setOfArcsRoad  \label{eq:KKTallRoad} \\
		&\left(\vTime\cdot  \traveltime_{ij} + \vDistS \cdot d_{ij}\right)  + \dualCust(j) - \dualCust(i) + \dualcS(i,j) \span\nonumber\\
		&+2\cdot\vQuadratic\cdot\flow{i}{j}^\star=0 & \forall m\in \setOfRequestsNumber, \arc\in\setOfArcsSubway \label{eq:KKTallSubway} \\
		&\left(\vTime\cdot  \traveltime_{ij}\right)  + \dualCust(j) - \dualCust(i)
		+2\cdot\vQuadratic\cdot\flow{i}{j}^\star = 0 \span\nonumber \\
		& & \forall m\in \setOfRequestsNumber, \arc\in\setOfArcsPedestrian \label{eq:KKTallPedestrian}\\
		&\left(\vDistR\cdot  d_{ij} +\vEnergy\cdot\energyR\right) + \dualVeh(j) - \dualVeh(i) + \dualcR(i,j)\span   \nonumber\\
		& +2\cdot\vQuadratic\cdot\flowReba{i}{j}^\star= 0
		&\forall \arc \in\setOfArcsRoad. \label{eq:KKTallRebs}
		\end{align}
	\end{subequations}
\end{small}
%\endgroup
	
	First, we prove that, if the KKT conditions for Problem~\eqref{eq:IAMoD} are satisfied, then the KKT conditions for each individual customer's routing Problem~\eqref{eq:customerproblem} are also  satisfied.
	
	The KKT conditions for customer $m$'s Problem~\eqref{eq:customerproblem} are
	
	\begin{subequations}
		\begin{align}
		&\vTime\cdot  \traveltime_{ij} +  \pOrigin{i} -  \pOrigin{j} +  \pRoad + \dualCustTilde(j) - \dualCustTilde(i) \span\nonumber\\ &+2\cdot\vQuadratic\cdot\flow{i}{j}^\star =0  &  \forall \arc\in\setOfArcsRoad \label{eq:KKTcustRoad}\\
		&\vTime\cdot  \traveltime_{ij} +  \pSubway + \dualCustTilde(j) - \dualCustTilde(i) \span\nonumber\\
		& +2\cdot\vQuadratic\cdot\flow{i}{j}^\star =0 & \forall \arc\in\setOfArcsSubway \label{eq:KKTcustSubway} \\
		&\vTime\cdot  \traveltime_{ij} + \dualCustTilde(j) - \dualCustTilde(i)\span\nonumber\\
		& +2\cdot\vQuadratic\cdot\flow{i}{j}^\star =0 & \forall \arc\in\setOfArcs\setminus(\setOfArcsRoad\cup \setOfArcsRoad), \label{eq:KKTcustPedestrian}
		\end{align} 
	\end{subequations}
	where $\dualCustTilde$ is the dual variable associated with constraint \eqref{eq:customerproblem}.
	Substituting the prices \eqref{eq:prices:subwaytolls}-\eqref{eq:prices:AMoDroad} in the equations above shows that, if Eq.~\eqref{eq:KKTallRoad}-\eqref{eq:KKTallPedestrian} are satisfied, then Eq.~\eqref{eq:KKTcustRoad}-\eqref{eq:KKTcustPedestrian} are also satisfied with $\dualCustTilde = \dualCust$. Thus, if  $\big\{\{\flow{\cdot}{\cdot}\}_m,\flowReba{\cdot}{\cdot}\big\}$ is the optimal solution to Problem \eqref{eq:IAMoD}, then $\flow{\cdot}{\cdot}$ is also the optimal solution to Problem \eqref{eq:customerproblem} for each individual customer $m$ for the given prices and tolls \eqref{eq:prices:subwaytolls}-\eqref{eq:prices:AMoDroad}.
	
	Next, we turn our attention to the  AMoD operator's Problem~\eqref{eq:operatorproblem}.
	The KKT conditions for Problem \eqref{eq:operatorproblem} are
	\begin{align}
	\vDistR\cdot d_{ij} +\vEnergy\cdot\energyR+  \tRoad + \dualVehTilde(j) - \dualVehTilde(i) \span\nonumber\\ &+2\cdot\vQuadratic\cdot\flowReba{i}{j}^\star = 0 &  \forall \arc\in\setOfArcsRoad,
	\label{eq:KKTrebRoad}
	\end{align}
	where $\dualVehTilde$ is the dual variable associated with Eq.~\eqref{eq:operatorproblem}.
	Substituting the road toll $\tRoad$ from Eq.~\eqref{eq:prices:roadtoll} shows that, if Eq.~\eqref{eq:KKTallRebs} is satisfied, then Eq.~\eqref{eq:KKTrebRoad} is also satisfied with $\dualVehTilde =  \dualVeh$. Thus, $\flowReba{\cdot}{\cdot}$ is also the optimal solution to the AMoD operator's Problem \eqref{eq:operatorproblem} for a given customer demand and given tolls.
	
	This concludes the proof.
\end{proof}
\end{appendix}
\fi

\end{document}